\documentclass[12pt,a4paper,titlepage]{article}

\usepackage{fancyhdr}
\usepackage{bm}
\usepackage{graphicx}
\usepackage{hyperref}
\usepackage{float}
\usepackage{setspace}
\usepackage{subfig}
\usepackage{amsmath}
\usepackage{amsfonts}
\usepackage{color}
\usepackage{hhline}

\def\orcidID#1{\unskip$^{[#1]}$}

\begin{document}
\title{\includegraphics[width=0.8\textwidth]{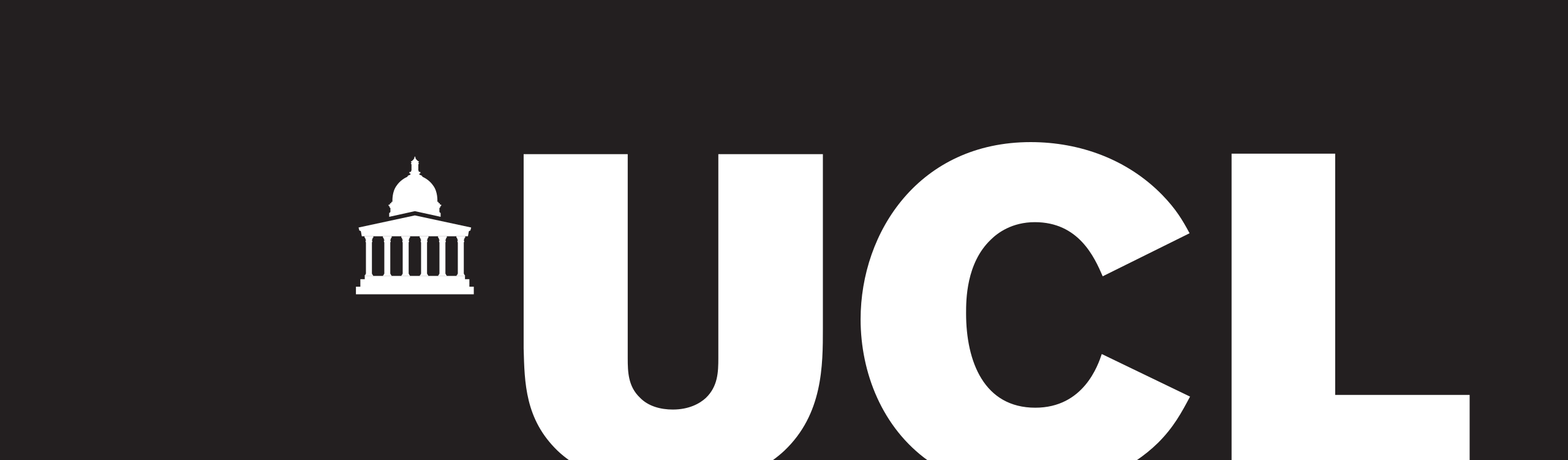}\\\phantom{}\\\textbf{Automated airway segmentation by learning graphical structure}
\author{Yihua Yang\orcidID{0000-0002-0860-7339}\\\phantom{}\\Master of Science\\\phantom{}\\Department of Physics \& Astronomy\\University College London\\Gower Street, London, WC1E 6BT\\\url{https://www.ucl.ac.uk}}}
\date{}

\maketitle

\newpage
\phantomsection
\addcontentsline{toc}{section}{Declaration}
\begin{spacing}{2}
{\Large I, Yihua Yang confirm that the work presented in this thesis is my own. Where information has been derived from other sources, I confirm that this has been indicated in the thesis.}
\end{spacing}
\newpage
\phantomsection
\addcontentsline{toc}{section}{Abstract}
\section*{Abstract}

\begin{spacing}{1.5}
In this research project, we put forward an advanced method for airway segmentation based on the existent convolutional neural network (CNN) and graph neural network (GNN). The method is originated from the vessel segmentation, but we ameliorate it and enable the novel model to perform better for datasets from computed tomography (CT) scans.
\begin{spacing}{2}
\end{spacing}
Current methods for airway segmentation are considering the regular grid only. No matter what the detailed model is, including the 3-dimensional CNN or 2-dimensional CNN in three directions, the overall graph structures are not taken into consideration. In our model, with the neighbourhoods of airway taken into account, the graph structure is incorporated and the segmentation of airways are improved compared with the traditional CNN methods.
\begin{spacing}{2}
\end{spacing}
We perform experiments on the chest CT scans, where the ground truth segmentation labels are produced manually. The proposed model shows that compared with the CNN-only method, the combination of CNN and GNN has a better performance in that the bronchi in the chest CT scans can be detected in most cases.
\begin{spacing}{2}
\end{spacing}
In addition, the model we propose has a wide extension since the architecture is also utilitarian in fulfilling similar aims in other datasets. Hence, the state-of-the-art model is of great significance and highly applicable in our daily lives.
\end{spacing}

\begin{spacing}{2}
\end{spacing}
{\bf Keywords:} Airway segmentation, Convolutional neural network, Graph neural network
\newpage

\phantomsection
\addcontentsline{toc}{section}{Acknowledgements}
\section*{Acknowledgements}
Throughout the research project, I have received precious guidance and support.
\begin{spacing}{2}
\end{spacing}
I would first like to thank my supervisors, Dr Joseph Jacob, Dr Eyjolfur Gudmundsson and Dr Moucheng Xu, whose expertise is invaluable for this research project. Without their guidance, I would not have achieved to complete this project.
\begin{spacing}{2}
\end{spacing}
In addition, I would like to appreciate to Dr Bridgette Cooper for her timely assistance during the whole term.
\begin{spacing}{2}
\end{spacing}
Finally, I am grateful to my parents. Their unconditional love for me motivates me to make progress.

\newpage
\tableofcontents
\listoftables
\listoffigures

\newpage
\pagenumbering{arabic}

\section{Introduction}
Trachea is one of the crucial organs in the human respiratory system. The diagnosis of many diseases such as idiopathic pulmonary fibrosis (IPF) depends on the observation of main trachea and bronchus.

In the contemporary society, computed tomography (CT) is one vital tool for the visualisation and evaluation of lung diseases \cite{jacob2016mortality}. Clinicians mainly rely on manual inspection of chest CTs to recognize the airways, which is time-consuming and even in some cases inaccurate. Therefore, the computer-assisted methods are necessary to be developed over the years to diagnose the lung fibrosis and measure the severity. For instance, axial CT scans are overlaid with different colours by CALIPER (Computer-Aided Lung Informatics for Pathology Evaluation and Rating software) in order to measure the IPF severity \cite{jacob2018predicting}. In this example, CALIPER is one tool for visualising and comparing reticulation of lungs. Besides, healthy human airway and airway with IPF differ in the airway branch structures \cite{smith2018human}. Clinicians are able to observe the tracheae from CT scans and give the possible diagnosis of patients.

We are aware of the significance of the computer-assisted techniques, so in this thesis, we are aimed to explore the segmentation of airways from chest CT scans, which is one of the key processes to the final automated diagnosis.

\begin{figure}[H]
\centering
\includegraphics[width=10cm]{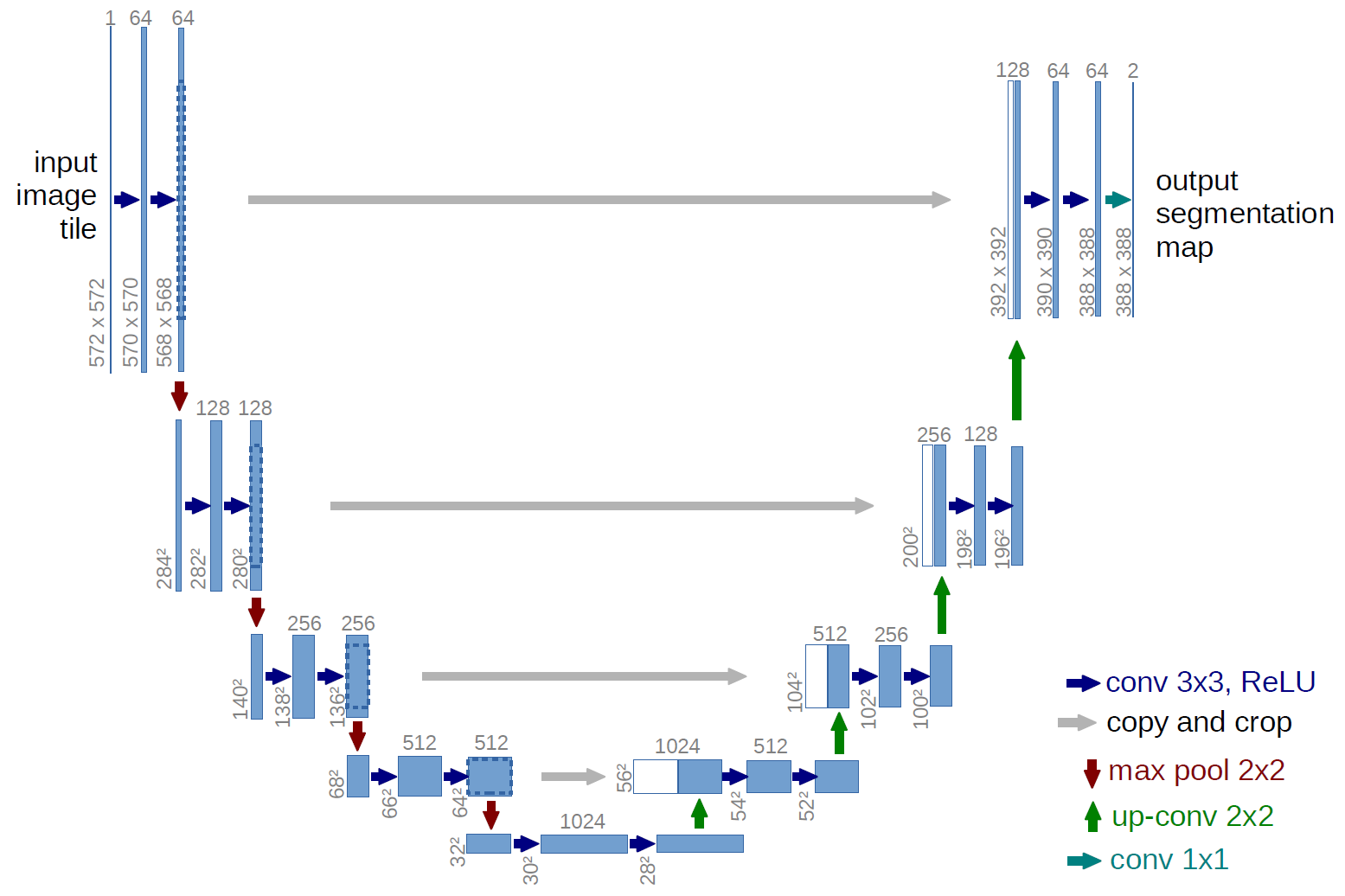}
\caption{U-net architecture \cite{ronneberger2015u}} \label{fig1}
\end{figure}

Currently, most popular methods related with image segmentation are based on probabilistic
point of view or deep neural networks. For the mean field network based graph refinement \cite{selvan2018mean}, mean field network is utilised in Markov random field to solve the computational burden problem. To be more specific, a subgraph from an over-complete input graph is obtained by a voxel classifier and Bayesian smoothing. For the deep learning methods, U-net is one popular model \cite{ronneberger2015u}, which is in the form of convolutional neural network (CNN). The model incorporates a contracting path and an expansive path (see
Figure~\ref{fig1}).

The U-net architecture is first designed especially for medical imaging \cite{ronneberger2015u} due to the good prediction of classification between tissues. We will imitate parts of the architecture and build our own network in the following sections.

\begin{figure}[H]
  \centering
  \subfloat[Segmentation]{\includegraphics[width=0.4\textwidth]{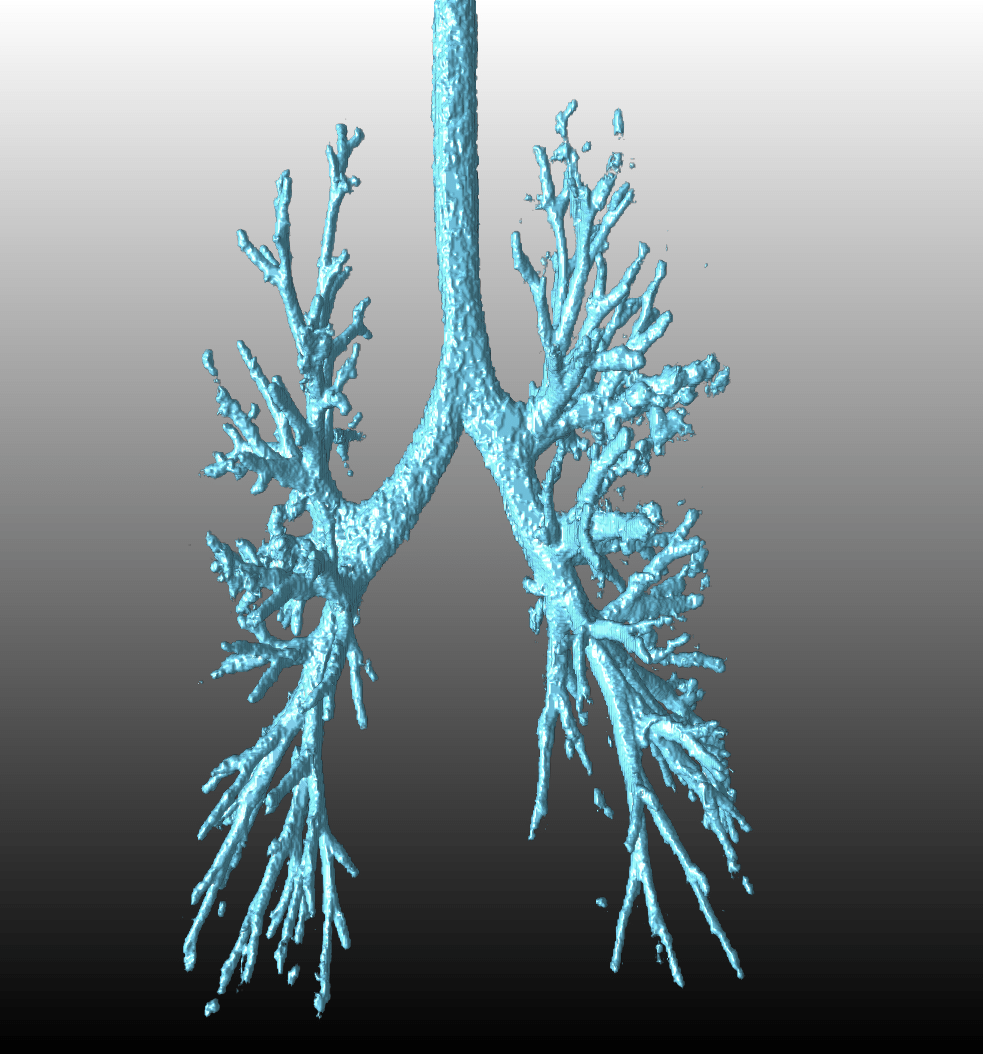}}
  \hfill
  \subfloat[Ground truth]{\includegraphics[width=0.4\textwidth]{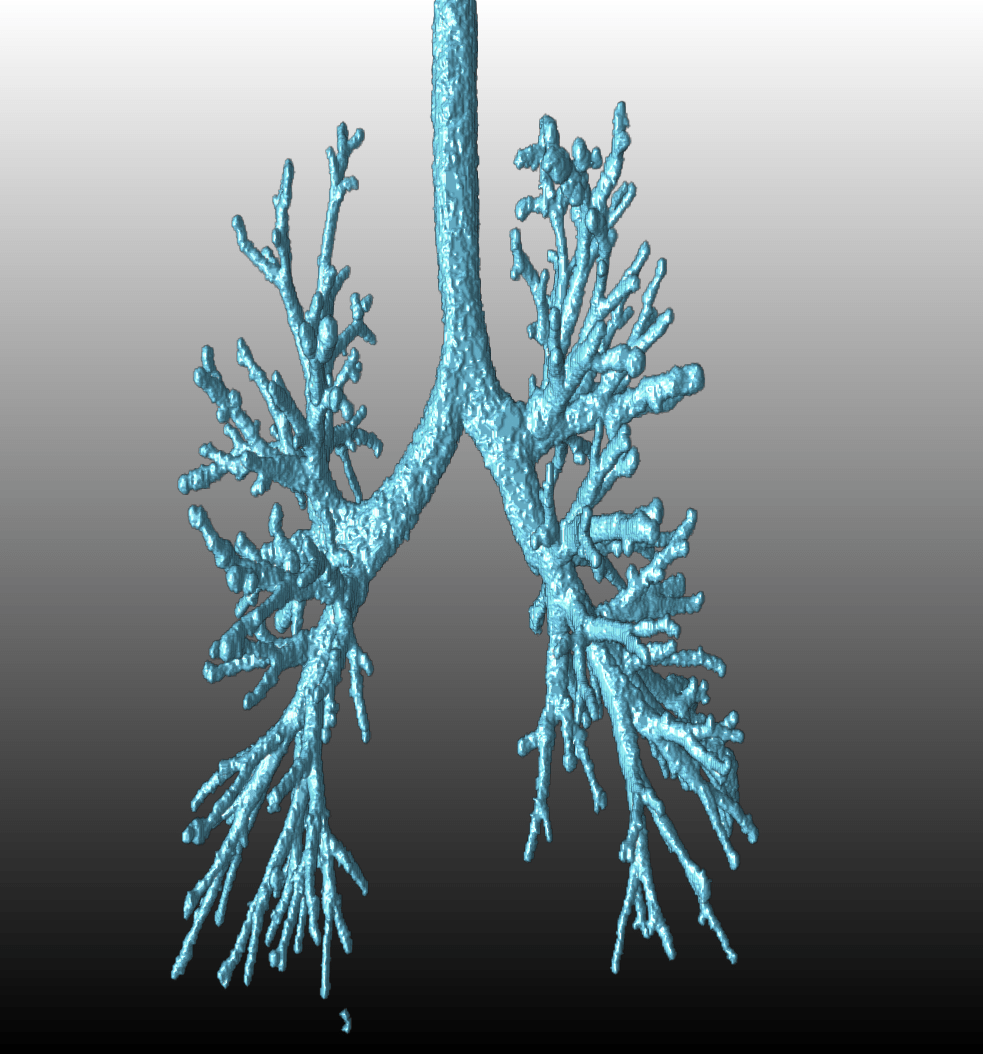}}
  \caption{Comparison of segmentation and ground truth \cite{juarez2018automatic}} \label{fig2}
\end{figure}

Based on the U-net architecture, the segmentation of airways has been done \cite{juarez2018automatic}. In this case, 3D U-net architecture is used while in the original architecture in the
Figure~\ref{fig1}, 2D U-net is built. The difference lies in the dimension of the intermediate layers. For 3D version, the depth is added in the 2D version so that the convolutional layers in 2D U-net is replaced by 3D convolutional layers. With cropping images, taking sliding-window approach and data augmentation process, the input images are put into the network and trained. Then the segmentation is successfully completed (see
Figure~\ref{fig2}).

The 3D version considers the airway structures of adjacent slices but the computational burden is heavier then 2D version.

The work of 2.5D convolutional network is a replacement of 3D version \cite{yun2019improvement}. It utilises 3 orthogonal directions to apply 2D convolutional layers to three adjacent slices and extract the features of airways through the same layers as U-net does. This method can reduce the computational burden without losing the 3-dimensional information.

Heretofore, all the networks are CNN-based methods. The weakness for CNN-based methods lies in the local appearances focused by the convolutional layers. They cannot provide precise segmentation even if the number of layers are increased \cite{maninis2016deep}.

In order to consider the global graph information, the graph neural network (GNN) is developed. The aim of GNN is to classify vertices on the graph, which is a irregular domain \cite{scarselli2008graph}. With GNN, the overall structure of the graph is learned and thus the features of the vertices in the graph are extracted.

With these state-of-the-art techniques, we can draw on the advantages of the several models and build our own Airway Graph Network (AGN).
\section{Prerequisites}
In this section, we give the fundamental knowledge of the whole model that we will build.
\subsection{Convolutional Neural Network}
For simplicity, we explain the convolutional layers in U-net architecture (see
Figure~\ref{fig1}). Besides, we give the definition of other layers and the training process.

A convolutional layer is acted as a function to generate convolved features \cite{indolia2018conceptual}. To be more specific, this algorithm receives an input image or any intermediate layers and output a series features captured by a kernel or filter containing parameters.
\begin{figure}[H]
\centering
\includegraphics[width=0.7\textwidth]{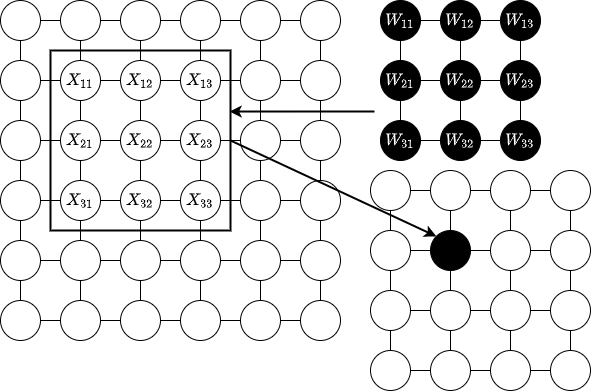}
\caption{Convolutional layer \cite{indolia2018conceptual}} \label{fig3}
\end{figure}
For instance, the input size is $6\times6$ and the kernal size is $3\times3$. With the stride size $1\times1$ and valid padding (padding = 0), we will have an output of size $4\times4$ (see
Figure~\ref{fig3}). The formula for the convolution operation is $a_{ij}=(W * X)_{ij}+b$, where $*$ is the element-wise multiplication, $a_{ij}$ is the output, $W$ is the kernel, $X$ is the input and $b$ is the bias term.

Usually, we will add a non-linear activation function called Rectified Linear Unit (ReLU) after each convolutional layer. $$f(x)=\max(0,x)$$ This function is used for several reasons. Firstly, if we do not employ non-linear activation function, then all the intermediate layers are only linear combinations of the initial input. Secondly, it is easy for calculating gradients when executing backpropagation. Last but not least, it filters the negative values and causes the sparsity of the network so that it can prevent network overfitting \cite{indolia2018conceptual}.

After some convolutional layers in the contracting path, they are followed by max-pooling layers. The max-pooling layers reduce the number of parameters that are trained by taking maximum values in the non-overlapping windows due to the fact that the size of the intermediate layers will be reduced so that the number of weights are reduced \cite{indolia2018conceptual}. In addition, they can avoid the sensitivity problem, which is aroused by the network's sensitivity in the location of features \cite{nagi2011max}. When the input is downsampled, the network is simplified with irrelevant elements being removed. Despite of the fact, it still allows the network to keep the large structural elements. In other words, the features that are required to be extracted are not lost.
\begin{figure}[H]
\centering
\includegraphics[width=0.7\textwidth]{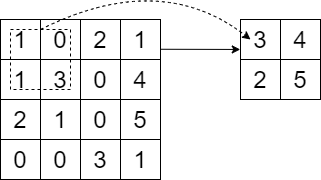}
\caption{Max-pooling operation \cite{indolia2018conceptual}} \label{fig4}
\end{figure}

For instance, for a $4\times4$ input and a window of size $2\times2$, the output will be $2\times2$ with maximum value in each window (see Figure~\ref{fig4}). 

\begin{figure}[H]
\centering
\includegraphics[width=0.7\textwidth]{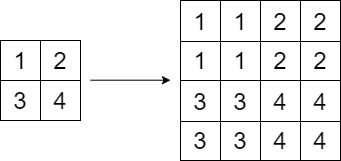}
\caption{Nearest-neighbour upsampling \cite{dumoulin2016guide}} \label{fig5}
\end{figure}

Another layer in the U-net architecture is the upsampling layer, which is the opposite to the max-pooling layer. The upsampling layer generates an output that has a dimension larger than that of the input. It is used to amplify the downsampled layers in the U-net architecture and keep the same size of the input since we are required to obtain the output of the same size as the input \cite{dumoulin2016guide}. The common algorithm is the nearest-neighbour method (see Figure~\ref{fig5}). It receives an element and copy it to the adjacent elements based on the times that are to be amplified.

When we upsample the intermediate layers in reality, we need to train the parameters to optimise the upsampling process. Hence, we use transpose convolution to not only upsample the layers, but also incorporate kernel parameters for training.

We observe from the contraction path to the expansive path, we also have to copy the intermediate layers to the upsampled layers. Since the convolution type is valid, it will not retain the same size. Hence, to ensure the same size, we use crop method to cut the layers into smaller size and concatenate them with the upsampled layers \cite{ronneberger2015u}.

All the layers in the U-net architecture have been introduced. Next process is the training process. With the forward passing, we can calculate all the neurons using weights that should be updated. In the last layer, we will have the final output based on these intermediate layers. The aim of this feedforward network is to calculate the loss of the model, but we need to optimise the weights in order to minimise the loss. The backpropagation process is all about calculating gradients of a chosen loss function with respect to weights of the model\cite{hecht1992theory}. With the chain rule, we can calculate all the gradients. Then, with the learning rate set, we can reduce the loss in the quickest direction, i.e.\ along with the negative gradient to ensure the maximum decrease. After numerous forward passing and backpropagation processes, we obtain a optimal model and we are able to evaluate the model.

With CNN, no matter what the exact model is, such as U-net, the networks perform similarly and output similar results, whose differences lie in the different characteristics of different datasets.
\subsection{Graph Neural Network}
Graph neural network (GNN) is different from CNN in that it considers the global graph structures \cite{scarselli2008graph}.

To begin with, we give the definition of a graph. A graph is composed of vertices and edges, denoted by $G=(V,E)$, where $V$ is the set of vertices and $E$ is the set of edges \cite{scarselli2008graph}.

Then, we give the definition of adjacency matrix. The adjacency matrix $A$ is used to represent the connection between vertices, which is composed of ones and zeros with one standing for joined vertices and zero for separate vertices \cite{scarselli2008graph}. Hence, we have $$A_{i,j}=\left\{\begin{array}{l}1, \text{if there is an edge between vertex } i \text{ and } j \text{ with } i\neq j, \\0, \text{otherwise.} \end{array} \right.$$
In the thesis, we only consider undirected graphs, so the adjacency matrix is symmetric.

With the vertices and edges, we define the features of each vertex. The features for vertex $i$ is denoted by $\vec{x_{i}}=(x_{i1},x_{i2},\cdots,x_{in})$, where $n$ is the number of features of the vertex $i$ \cite{scarselli2008graph}. Overall, in one graph $G$ which contains $N$ vertices and $n$ features for each vertex, the feature matrix is denoted as $X=(\vec{x_{1}},\vec{x_{2}},\cdots,\vec{x_{N}})^{\bm{T}}$.

There are many variants of GNN. In the thesis, we consider graph attention networks (GAT).

GAT is an attention-based architecture that attends adjacent vertices of one vertex and computes the hidden representations of it \cite{velivckovic2017graph}. In the followings, we give the processes of how GAT works.

In one graph attentional layer, the input is a set of vertex features, which is denoted by $X=(\vec{x_{1}},\vec{x_{2}},\cdots,\vec{x_{N}})^{\bm{T}}$. $N$ is the number of vertices and in each vertex $\vec{x_{i}}=(x_{i1},x_{i2},\cdots,x_{in})$, where we assume there are $n$ features in each vertex \cite{velivckovic2017graph}. The output is a set of new features. The number of new features in each vertex is assumed to be $n'$.

In the first step, the features in each vertex are applied to a linear transformation by a weighted matrix $W$. Namely, $\vec{z_{i}}=\vec{x_{i}}W$ \cite{velivckovic2017graph}. Then, we calculate the attention coefficients, which determines the importance of adjacent vertices. The formula is given by $e_{ij}=\text{LeakyReLU}\left(\left(\vec{z_{i}}\| \vec{z_{j}}\right)\vec{a}\right)$, where $a$ is the shared attentional mechanism that can be learned, $\|$ is the concatenation operation and $\text{LeakyReLU}(x)=\left\{
\begin{aligned}
x &, \text{ if } x>0, \\
0.01x &, \text{ otherwise.}
\end{aligned}
\right.$ Afterwards, we normalise the attention coefficients to ensure the common scaling across all vertices by means of the Softmax function \cite{velivckovic2017graph}, where $\text{Softmax}(e_{ij})=\frac{\exp(e_{ij})}{\sum_{k\in \mathcal{N}_{i}} \exp(e_{ik})}$ with $k\in \mathcal{N}_{i}$ meaning some vertex is in the set of adjacent vertices of vertices $i$. The normalised attention coefficients are denoted as $\alpha_{ij}=\text{Softmax}(e_{ij})$. Finally, we compute the new features using formula $\vec{x_{i}}'=\text{ELU}(\sum_{j\in \mathcal{N}_{i}} \alpha_{ij}z_{j})$, where $\text{ELU}(x)=\left\{
\begin{aligned}
x &, \text{if } x>0, \\
a(\exp(x)-1) &, \text{otherwise}
\end{aligned}
\right.$
with $a$ the given parameter.

We can also apply different attentional mechanisms to generate new features. In other words, we can initialise different mechanisms $\vec{a}$ and apply them to each pair of adjacent vertices to generate unnormalised attention scores \cite{velivckovic2017graph}. In this case, we just concatenate all generated new feature vectors and obtain the final matrix of new features all vertices.

In addition, if we use the features as the final layer in the model, we will transform them to a probability map \cite{velivckovic2017graph}. To this end, we will not use $\text{ELU}$ function to obtain new features. Instead, we obtain all set of new features and take the average of them. Lastly, we use a sigmoid function to transform the features to probabilities, where $\text{Sigmoid}(x)=\frac{1}{1+\exp(-x)}$. Namely, $\vec{x_{i}}'= \text{Sigmoid}(\frac{1}{K}\sum_{k=1}^{K} \sum_{j\in \mathcal{N}_{i}} \alpha_{ij}z_{j})$, where $K$ is the number of attentional mechanisms.
\begin{figure}[H]
\centering
\includegraphics[width=0.7\textwidth]{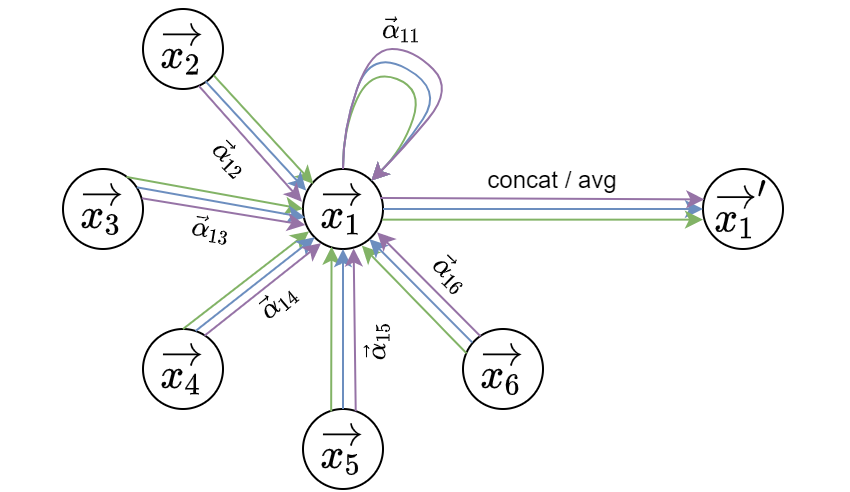}
\caption{Illustration of different attentional mechanisms (with K = 3) in GAT \cite{velivckovic2017graph}} \label{fig6}
\end{figure}

As shown in the Figure~\ref{fig6}, different attentional mechanisms are applied to generate the final new features or probabilities. For instance, if the number of input features is $n$ and the number of attentional mechanisms is 3, then the counterparts of the output features is $n'=3n$ with concatenation.
\subsection{Vessel Graph Network}
\begin{figure}[H]
\centering
\includegraphics[width=0.7\textwidth]{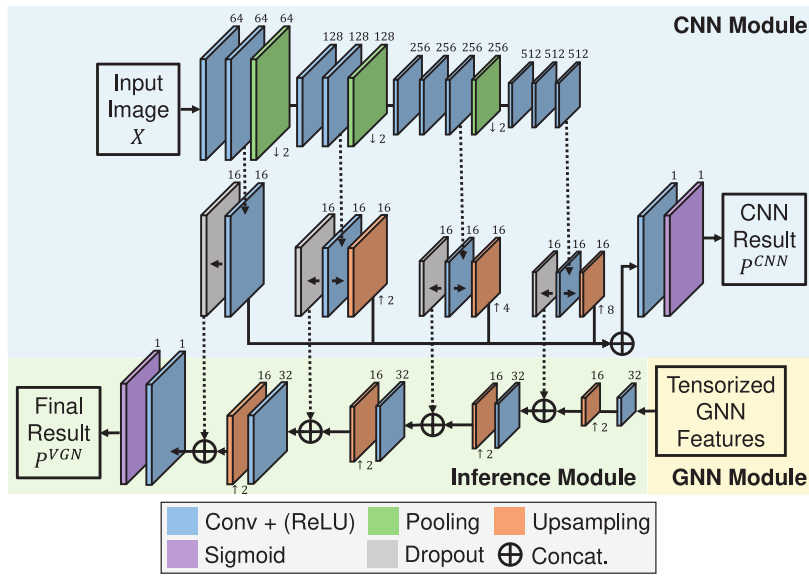}
\caption{Vessel graph network \cite{shin2019deep}}\label{fig7}
\end{figure}
With the introduction of CNN and GAT, we introduce a novel model called Vessel Graph Network (VGN) \cite{shin2019deep}. As a matter of fact, the model is a combination of CNN and GAT, which enables the network to refine the segmentation work. We will build our own AGN referring to the overall structure of VGN.

It is shown in the Figure~\ref{fig7} that there are in general three modules: CNN module, GNN module and inference module.

In the CNN module, the general concept is the CNN in the contracting path of the U-net architecture (see
Figure~\ref{fig1}). Different from the U-net, the module refers to the Deep Retinal Image Understanding network (DRIU) \cite{maninis2016deep} based on the model built by Visual Geometry Group (VGG) with depth of 16 (VGG16) \cite{simonyan2014very}.

On one hand, similar to the U-net architecture, it learns features on the regular grid and output a probability map of the same size as the input with only one channel instead of three in the input \cite{shin2019deep}. To be more specific, with the 3-channel input, which is a color image, we aim to classify if each pixel value in the image is the target we need to segment. On the other hand, we no longer use the copy and crop processes. Instead, we use another convolutional layer to decrease the number of channels in each intermediate layer after the convolution and max pooling pairs. Afterwards, instead of using queue-style upsampling method in the U-net architecture (see
Figure~\ref{fig1}), we upsample each of the convolved layers with different times and concatenate them together. The difference in the DRIU model \cite{maninis2016deep} has several advantages. Since the input images contain many features, in each convolution and maxpooling pairs we extract some and reduce the number of channels to save memories. In addition, after we obtain multi-scale features, we concatenate them together to reflect all the extracted features. This is more effective than the U-net architecture in that the expansive path illustrates a superimposed process. In other words, the extracted features in each convolution and max pooling pairs are likely to be lost when they are updated and entered into the next upsampling process. Also, U-net architecture is less effective due to the fact that there are still covolution layers after each upsampling layer. Hence, with the output, we are able to obtain an approximate segmentation.

The CNN module will be pretrained and the result will be utilised in the GNN module \cite{shin2019deep}.

Subsequently, in the GNN module, the features in the regular grid will be regarded as in the irregular grid that is a graph and learnt through GAT \cite{shin2019deep}. To be more specific, for each vertex in the graph, instead of learning local features, GAT learns hidden representations between adjacent vertices. The influence of adjacent vertices is considered so that the global graph features will be learnt. For instance, through GNN module, if one vertex $A$ is considered to be adjacent to another vertex $B$ that is much likely to be the target, i.e.\ the probability for the adjacent vertex is large from CNN module, then this vertex $A$ will be much probable to be the target based on the connectivity in the graph \cite{shin2019deep}. This function cannot be realised in the CNN module so that we utilise the GNN module in order to refine the recognition of vertices that cannot be detected in the CNN module.

With the input from the pretrained CNN module, a set of new features are output from the GNN module, which contains global graph features, compensate for the coarseness in the CNN module. The output of GNN module will be regarded as the input to the final inference module \cite{shin2019deep}.

Eventually, in the inference module, we follow the similar processes in the expansive path in the U-net architecture (see
Figure~\ref{fig1}). The aim of the module is to integrate the features extracted from CNN module and GNN module \cite{shin2019deep}. With the input of graph features, we utilise the convolution and upsampling pairs to adjust the size of the layers as well as incorporating features in CNN modules, which are represented as several intermediate layers in the Figure~\ref{fig7}. There is a drouout technique in the inference module that is distinguished from the U-net architecture. In this scenario, this dropout regularisation method randomly zeros some values in the intermediate layers in the CNN module based on the set probability. It can prevent the overfitting of the model and reduce the influence of CNN module that is possible to cover up the features in the GNN module \cite{shin2019deep}. After the same processes of convolution, upsampling and concatenation of the dropout, we will obtain a result of probability map that illustrates whether the pixel in the image belongs to the target or not.
\begin{figure}[H]
\centering
\includegraphics[width=\textwidth]{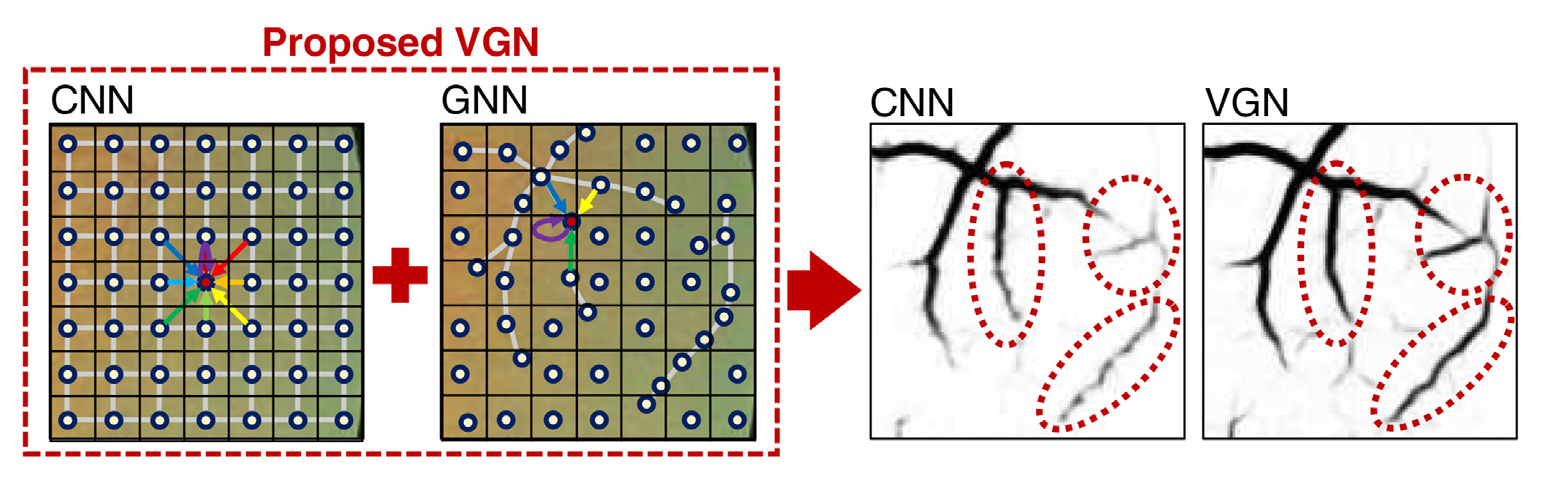}
\caption{Refinement of the VGN \cite{shin2019deep}}\label{fig8}
\end{figure}

In the Figure~\ref{fig8}, the VGN illustrates the refinement of the target, which in this case is the vessels. With GNN module, the vertices sampled in the same vessel are connected to each other. With the feature recognised, the whole model is able to perform better than the common CNN models \cite{shin2019deep}.

We have given the brief description of all the prerequisites before we give our own AGN. In the next section, we will define AGN and give detailed explanation of the network together with the dataset,the training parameters and training processes, the results and the discussions.
\section{Airway Graph Network}
\begin{figure}[H]
\centering
\includegraphics[width=\textwidth]{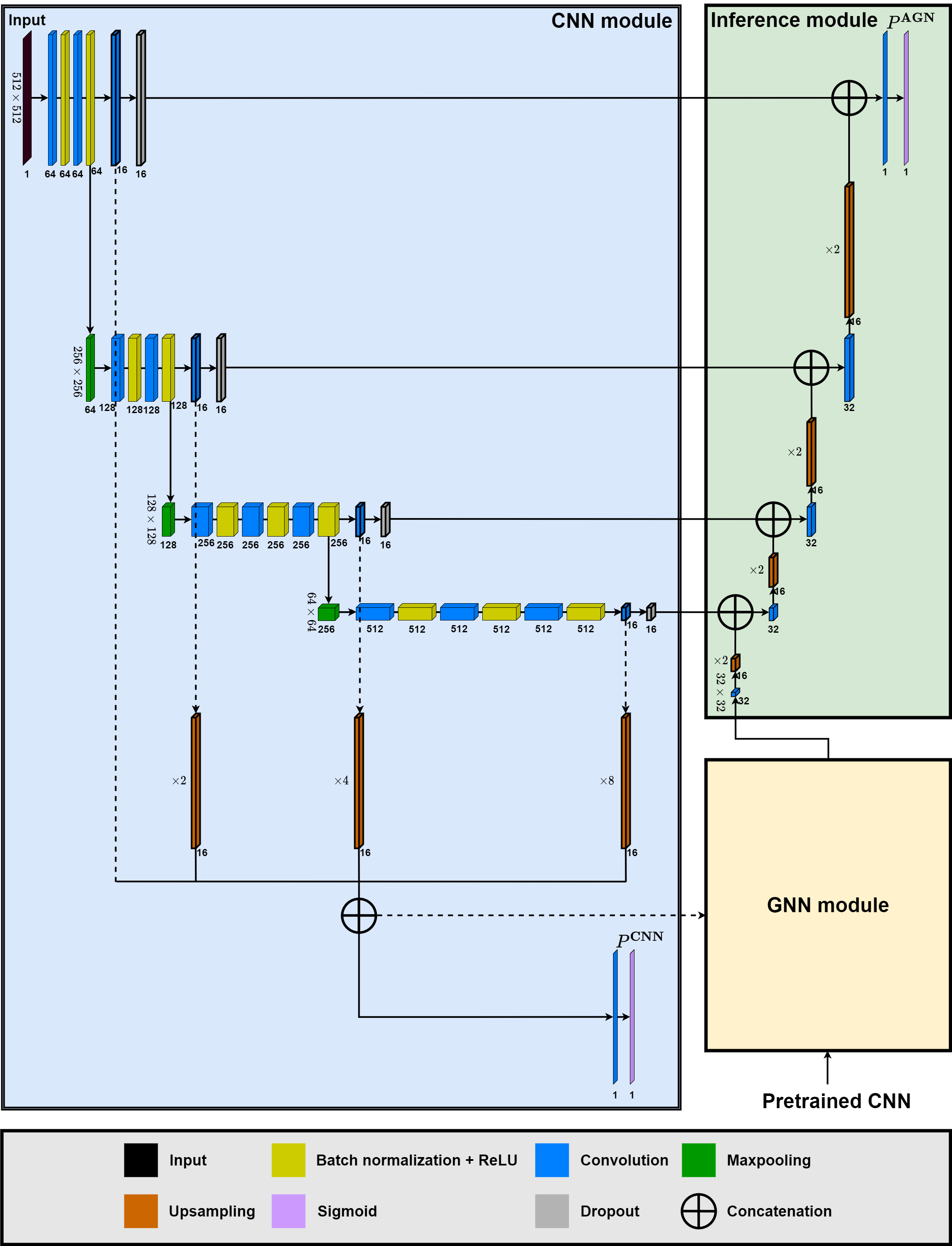}
\caption{Detailed architecture of the AGN}\label{fig9}
\end{figure}

In this section, we build our own model AGN. First and foremost, we give the detailed AGN architecture in the Figure~\ref{fig9}.

\subsection{Overall architecture}
In our model AGN, there are three modules in total: CNN module, GNN module and Inference module, which are similar to the VGN (see
Figure~\ref{fig7}).

In the CNN module, we input slices of CT scans and output probability maps, which indicates an approximate segmentation of the airways. The aim of this module is to find signs of airways in the slices of the CT scans. However, this module cannot segment the exact airways, especially for the bronchi. Similar to the VGN, the CNN module in the AGN learns features in the regular grid \cite{shin2019deep}. To be more specific, for an input slice, the pixel values are assessed by the convolutional layers with max-pooling layers. For the light pixels, it is possible for the convolutions to distinguish tracheal cartilages from other tissues since the nearby pixel values show significant differences between each other. However, not all light pixel values with these features can be classified as targets. Due to the fact that the convolution considers the local structures, the pixel values in the specific positions in the slices will be considered as potential targets. In other words, other parts such as aortas will not be segmented erroneously. Also, with the max-pooling layers, the CNN module enables the model to highlight the prominent low-level features. After obtaining all the  intermediate convolutional layers, the upsampling procedures are executed and each layer is resized into the original size of slices. However, the CNN module is limited in performance due to its focal point on local appearance.

In the GNN module, we input the original CT slices and output a set of new features based on the specific graphs we build, which indicates the connectivity between vertices in the graphs. In other words, the GNN module acts like the CNN module but the difference lies in how the features are extracted. More precisely, The new features reflect global structures, which are equivalent of local features in the CNN module. The input goes through all the layers in the CNN module before the final convolution to obtain the original features. Together the vertices chosen in the graph, we use GAT to generate a set of new features that represent whether the vertices belong to the same airway or not. It is effective for the bronchi since the structure of tube reveals the curve-shaped light pixel values in the input.

In the Inference module, we utilise the expansive path in the U-net architecture (see
Figure~\ref{fig1}) so that the features in both CNN module and GNN module can be combined. Similar to the VGN (see Figure~\ref{fig7}), we abandon the copy and crop. Instead, we employ convolutions to first reduce the sizes and then upsample to amplify the corresponding times. The module receives the intermediate layers in the CNN modules and the new features in the GNN modules. It outputs the final probability maps that predict the airways.

To sum up, the segmentation work is just the pixel-wise classification work. It combines the useful features  both in local and global appearances.

\subsection{CNN module}
We employ the DRIU network \cite{maninis2016deep} in the CNN module. Note that this network is also based on the contracting path of the U-net architecture (see Figure~\ref{fig1}).

The main layers that extract the features in this module are the convolutional layers. There are 15 convolutional layers in total. With the input slices of CT scans, we first utilise a convolutional layer to expand the number of channel from 1 to 64, which is the common manipulation in almost every existing network. After that, we apply a different technique between AGN and VGN or U-net. In both of the VGN and U-net architecture, the following layers are the ReLU layers. However, in our case, due to the particularity of the input, which are the slices of CT scans, we have to utilise a batch normalization technique \cite{ioffe2015batch}. To be more specific, for the input slices, the first convoluional layer extracts the features and reflect the features by transforming the input to with initialised weights. As usual, the followed ReLU activation function, which in our case is a layer, should filter out the negative values and retain the useful neurons. However, since the input slices are composed of negative hundreds or thousands to hundreds or thousands values, which make up the tensors, even if we rescale them by adjusting the levels and windows of the slices, when passing through the ReLU layer during training, the features extracted by the convolutional layer will be lost. Actually, the negative values should be retained but they are lost in training. In other words, the network will stop training in the early stages while for other datasets such as the 3-channel color images, the problem will not occur. By changing the mean and variance of the values in the tensor to almost zero mean and almost unit variance, the features extracted but reflected as negative hundreds or thousands are retained. The problem caused by the particularity of CT slices is tackled. In addition, there are several other advantages for the batch normalization process. Without batch normalization, there is an internal covariate shift between layers, then the distribution of the features or the outputs changes. It forces the following layers to adapt it so that the learning speed is slow \cite{ioffe2015batch}. When applying batch normalization, higher learning rates are accepted. Hence, the training of the network is faster and more efficient. Anyway, the aim of batch normalization is to prevent the invalidation of our model.

The next layer is the ReLU layer, whose function has been introduced and is the same as that in the U-net architecture or the VGN. Subsequently, the CNN module goes deeper and repeats the convolution, batch normalization, ReLU pair again like in all other models. Then, there are two routes for the intermediate tensors. On one hand, we utilise another convolutional layer to reduce the size of the tensors into quarter. This step is aimed to be prepared to execute the final concatenation of all upsampled result. On the other hand, we do the maxpooling. There are 3 max-pooling layers in the CNN module in total. As for the number of maxpooling layers, we consider the size of the input slices. Compared with CHASE\_DB1 dataset in the VGN \cite{shin2019deep}, the size of the input slices is smaller, so we do not employ more maxpooling and upsampling. Like in the U-net architecture (see Figure~\ref{fig1}), we reduce the size of the tensors in the maxpooling to extract sharper features in the tensors.

Afterwards, we repeat the convolution, batch normalization, ReLU pairs in two times and once more execute the quarter-sized convolution and maxpooling respectively. 

With the above whole processes repeated in 4 times in total, we obtain 4 quarter-sized tensor in channel with different size. Eventually, we upsample each of them with 1 time, 2 times, 4 times and 8 times respectively and concatenate them together to derive the original-sized tensor with 64 channels. This process combines each set of features extracted in different depth of the CNN module.

With the final convolution to transform the multi-channel tensor to 1-channel and through a sigmoid activation function, we obtain the probability map $P^{\text{CNN}}$, which consists of pixel-wise probabilities. The reason for the final convolution not followed by batch normalization and ReLU is due to its final layer position. The one-channel tensor has already contained the combined features implicitly and there is no other convolutional layers  following so that the manipulation of the batch normalization and ReLU layer is redundant. This probability map $P^{\text{CNN}}$ is coarse and will be refined by the GNN module.
\subsection{GNN module}
We employ the GAT \cite{velivckovic2017graph} in the GNN module. For the GAT, it is vital to prepare the input so that we can execute the GAT. Hence, the GNN module is actually composed of two procedures: building graphs and executing the GAT.
\subsubsection{Vertex sampling}
There are two inputs for the GAT, one being the original features and the other being the adjacency matrix. The original features requires the specific vertices to be selected and the adjacency matrix should be built based on the original features.

To begin with, we will obtain a pretrained CNN module that can provide the probability map given that the input slice is passed through the CNN module. Then, for each input $X$, the probability map $P^{\text{CNN}}$ is of the original size. We have to split $P^{\text{CNN}}$ to obtain the chosen vertices. Hence, we use $\delta$ to control the number of grids in each probability map. To be more precise, assuming the size of the input slices is $h\times w$, we split the original size into grids of size $\left\lceil \frac{h}{2^{\delta}} \right\rceil \times\left\lceil \frac{w}{2^{\delta}} \right\rceil$, where the $\left\lceil \text{ } \right\rceil$ gives the integer that rounds up. Also, these parts do not overlap. For each grid of the size $2^{\delta}\times 2^{\delta}$, we select one vertex that represents the lightest pixel. The light pixels in the grid tend to represent tissues such as heart and lungs instead of air or background. Hence, it is essential for us to pick them up and distinguish them with each other. As a matter of facr, there are several possibilities for this chosen pixel. For one case, if for all the pixels in the grid the values are the same, then the center of the grid is selected, representing the background. For another, if not all the values are the same, then there exists a maximum value or some maximum values. On one hand, if the number of maximum value is one, then we just select it. On the other hand, if there are several maximum values, then we take a random one. After this procedure, there are $\left\lceil \frac{h}{2^{\delta}} \right\rceil \times\left\lceil \frac{w}{2^{\delta}} \right\rceil$ number of vertices selected and we are able to do the edge construction based on the vertices.

To sum up, the vertex sampling procedure inputs probability maps and select vertices to build graphs for the GAT. Note that we have to save the positions of the selections, so the number $\delta$ is important for the construction. If the delta is too small, the network will suffer from memory insufficiency. However, if the $\delta$ is too large, then the graph we build is coarse and it is likely to miss important vertices even if the relationship between vertices that are distant from each other may be built.
\begin{figure}[H]
\centering
\includegraphics[width=0.5\textwidth]{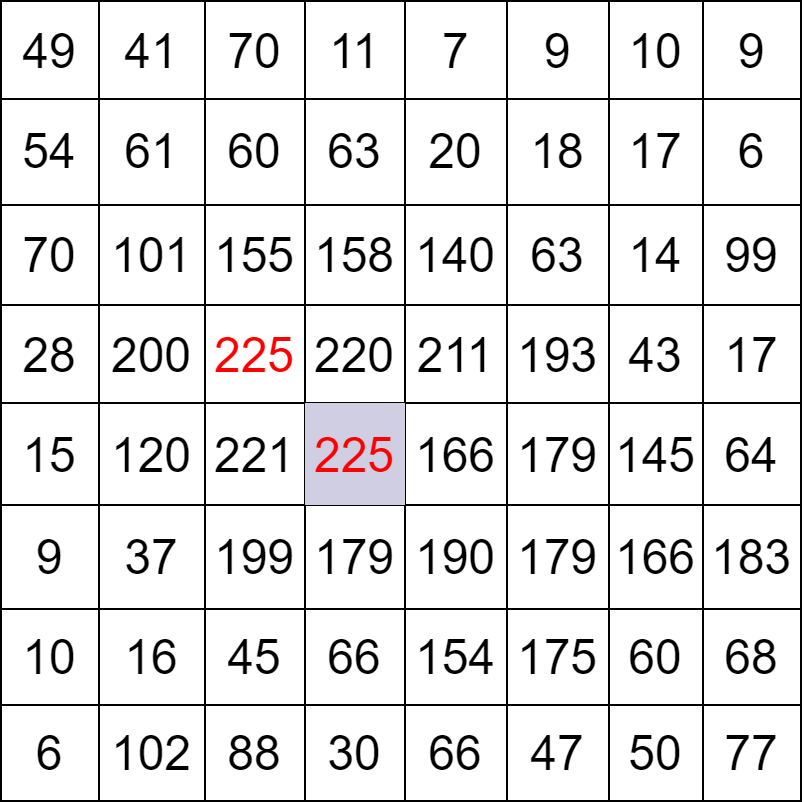}
\caption{Vertex sampling}\label{fig10}
\end{figure}
In the Figure~\ref{fig10}, for instance if the $\delta=3$, then there are 64 pixels in each grid with values ranging from 0 to 255. We randomly pick up one lightest vertex in this grid and keep record of the position in the global probability map.
\subsubsection{Edge construction}
The other input for the GAT is the adjacency matrix. Currently, we have already obtained the selected vertices. In the next step we utilise the positions of the vertices to attain the original features since we do not need the probabilities.

In the architecture of CNN module (see
Figure~\ref{fig9}), there is a concatenation before the final convolutional layer. It combines different features in different depths of the network so that we utilise this intermediate layer and consider the concatenated tensor as the original features.

With the selected vertices, we take the corresponding positions in this original feature map for every input probability map and delete other unselected elements. Then the original feature map will become a feature map with a reduced size $\left\lceil \frac{h}{2^{\delta}} \right\rceil \times\left\lceil \frac{w}{2^{\delta}} \right\rceil$.
\begin{figure}[H]
\centering
\includegraphics[width=\textwidth]{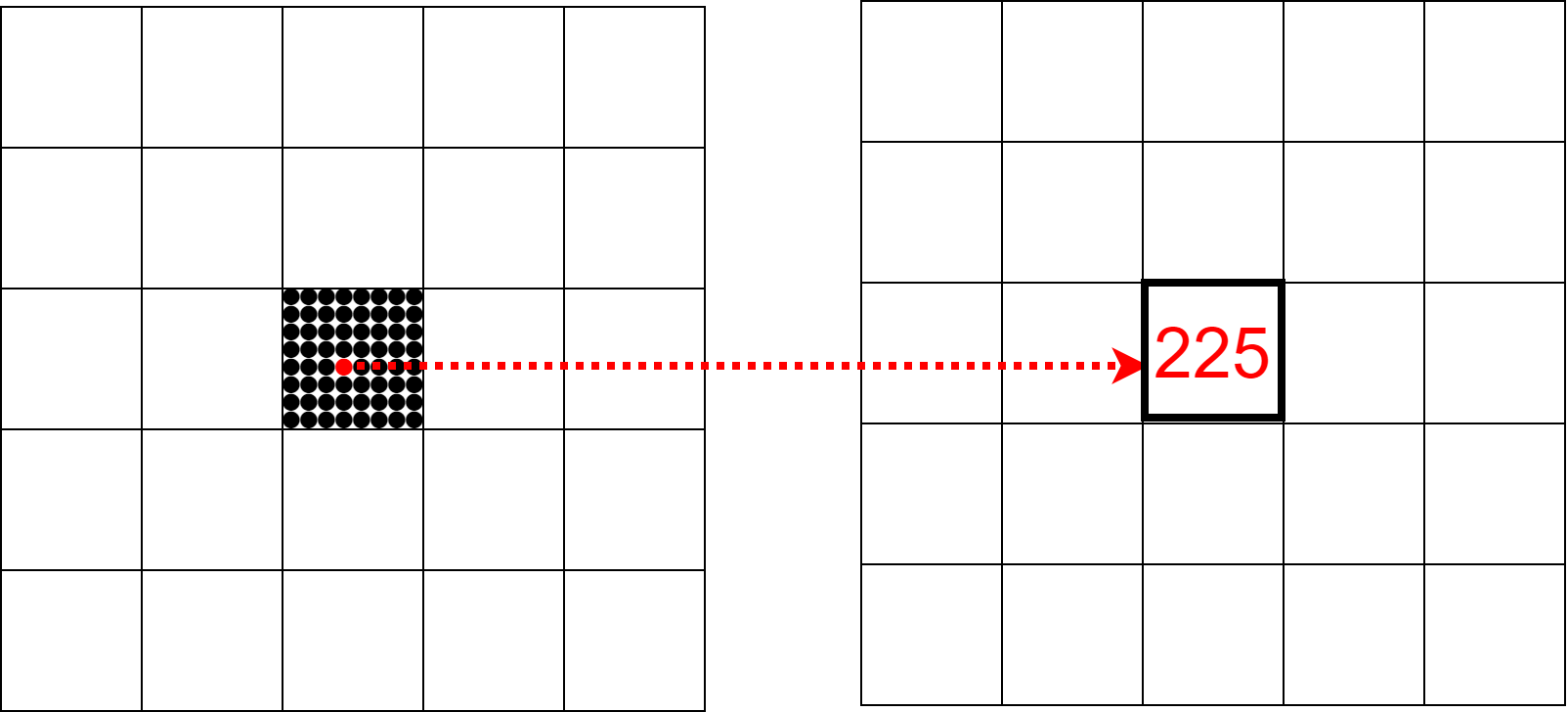}
\caption{Feature map of reduced size}\label{fig11}
\end{figure}
For instance, in the Figure~\ref{fig11}, each grid is of the size $8\times 8$ and the red dot represents the selected vertex of the maximum value 225 in the vertex sampling procedure, then we delete all other elements in that grid. We repeat the process and obtain a feature map of the reduced size. 

After we obtain the reduced feature map for the GAT, we have to obtain the adjacency matrix. In order for it, we have to build edges which are undirected in our case. We consider both of the difference between the probabilities and the Euclidean distance in the probability map of original size to decide whether we assign an edge or not. The method we apply is considering one measure called geodesic distance \cite{shin2019deep}. The distance is defined by calculating all the cumulative sums of probability differences in each path of two elements. There are many zigzag paths of connecting two elements and our aim is to find out the smallest one. If the value is less than a threshold $d$, then we will assign an edge between the vertices.

The realization of the method is difficult due to the possibilities of many paths. Hence, we apply a replacement. The scikit-fmm package \cite{furtney2019scikit}, which is the abbreviation for scientific Python toolkit fast marching method, contains a function of 'travel\_time'. It receives the array, which contains a zero contour and the interface propagation speed and returns the time spent. For instance, for one probability map, we create a new array that contains only ones except for the position for one selected vertex. That position is assigned negative one. Hence, there exists a zero contour for the array. We use the 'travel\_time' function and obtain the geodesic distance for every vertex. Finally, we compare the values for the selected vertices in the corresponding positions and assign the edge if the values are less than the threshold $d$.

We repeat the process for each vertex and can obtain the adjacency matrix.

Now that we have obtain the original features and adjancency matrices, we will pass them through the GAT and obtain sets of new features. In our scenario, we apply 4 different attentional mechanisms with the output features being 16. Hence, after concatenation, the number of features will be 64. The number of input features are 64 and they are within each channel for each pixel. Note that there is an extra process of flattening the features into rows and put the new features back into the grids since the GAT deals with row-wise features.
\begin{figure}[H]
\centering
\includegraphics[width=\textwidth]{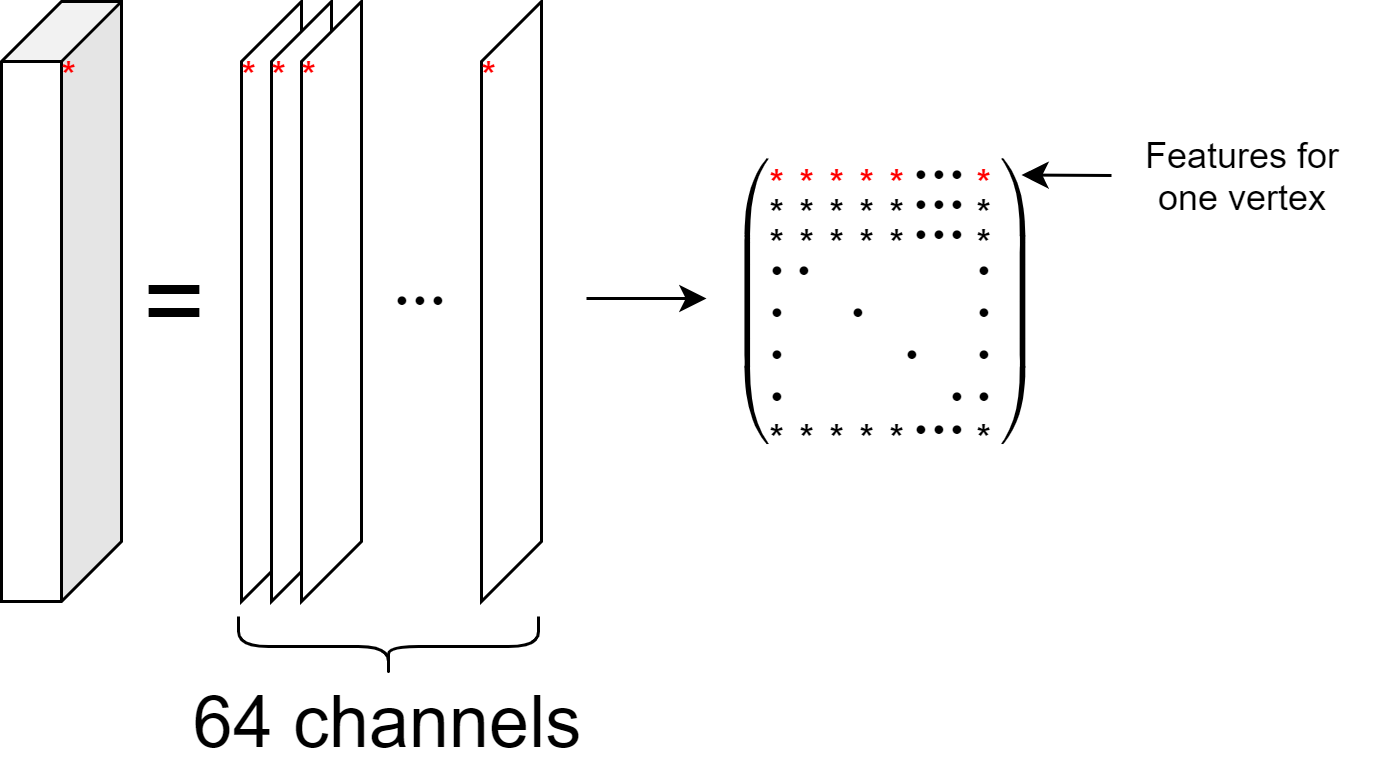}
\caption{Flattening of the features}\label{fig12}
\end{figure}
For instance, in the Figure~\ref{fig12}, it illustrates how selected vertices are flattened and put into each row. The number of rows are the number of vertices and the number of columns are 64.
\subsection{Inference module}
In the Inference module, the inputs are the new features extracted from the GNN module, which represent the graph structures including the selected vertices and the undirected edges. The module is aimed to combine the features in the CNN module, i.e.\ the local features should be combined as well. Hence, this module imitates the expansive path of the U-net architecture (see
Figure~\ref{fig1}).

To be more specific, after we obtain the new features lied in the rows. We put them back into channels and each vertex is put back into a regular grid. It is just the reverse of the process in the Figure~\ref{fig12}. Afterwards, we do a pair of convolution and upsampling pair, which doubles the size of the tensor and adjusts the number of channels to 16 for the concatenation of the previous CNN module. Before concatenation, the technique of dropout layers are applied similar to the VGN \cite{shin2019deep}. Four such pairs are executed and in the final concatenation, the tensor of the original size and number of channels being 32 are obtained. In the penultimate convolutional layer, the number of channels are reduced to one, which combines all the implicit features. Lastly, we utilise a sigmoid activation to transform the features into probabilities, which is the final prediction.

During training or evaluation, the single input CT slice is passed through all three modules sequentially to generate the final result.
\section{Experiment and Results}
In this section, we will utilise the model we build to experiment on the dataset we prepare. In addition, we will give the description of the codes we use and illustrate the result after executing the codes. Last but not least, we will discuss on the model itself and list the shortcomings and put forward the potential improvements on them.
\subsection{Dataset}
The proposed AGN model is trained and tested on the dataset of chest CT scans from nine patients who suffer from IPF.

Firstly, we have to do the segmentation manually to ascertain the ground truth labels, i.e.\ the airways in these CT scans. We employ the software '3D Slicer' to add the data of the original 3D chest CT scans for each patient. Then we initialise the segmentation by the built-in tool 'Grow from seeds'. To be more precise, we create two segmentation labels called 'Lung' and 'Trachea' in the 'Segmentation Editor' module. Then for each plane (overall axial, coronal, sagittal) of the CT scans we select some slices and paint the corresponding lung and tracheae under the corresponding labels. We repeat the process in three consecutive slices. Eventually, we utilise the 'Grow from seeds' tool to segment the airways automatically. However, the segmentation is coarse and is required to be refined. Hence, we paint the bronchi manually pixel by pixel in each slice. The distance between each slice is one millimetre and we obtain hundreds of slices for each CT scan.

After we finish the improved segmentation manually, we have to take slices of the CT scans and segmentations. We determine to employ the axial plane. Therefore, for each patient, we output two files, one being the slices of CT scans and the other being the segmentation labels.

For the simple access of these data, we integrate the CT scans and segmentations into two arrays of the same size. That is to say, two slices from the two arrays in the same position correspondingly represent the slice for the CT scan and the segmentation for it respectively.

Finally, we obtain the two arrays of the same size: (512, 512, 2566), where 2566 is the number of slices and 512 is the height and width for each slice.
\subsection{Evaluation Details}
After we obtain the dataset with easy access, we have to train the dataset. Since the number of slices is large, we will reduce the size of slices by deleting the useless slices. Due to the fact that not every slice contains airways, we filter out these which does not contain airways, i.e.\ the corresponding slice in the second segmentation array is all zero. Then we obtain the two arrays of the same size: (512, 512, 1964), where the only first slice does not contain any airway, but it affects nothing. Compared with the number of images to be trained in the VGN model, which are the retinal vessel images, the size of the CT slices is smaller and the number of them are larger, which are revealed in the Table~\ref{tab1}.

\begin{table}[H]
\centering
\caption{Datasets details}\label{tab1}
\begin{tabular}{lll}
\hline
Dataset & Airway & CHASE\_DB1 \\
\hhline{===}
Number of inputs & 1964 & 28\\
(train/test) &  (1500/464) & (20/8)\\
Resolution & $512\times 512$ & $960\times 999$\\
\hline
\end{tabular}
\end{table}

The ratio of the training data to all the data is similar in both datasets, which is around 70\%.

\begin{figure}[H]
\centering
\includegraphics[width=\textwidth]{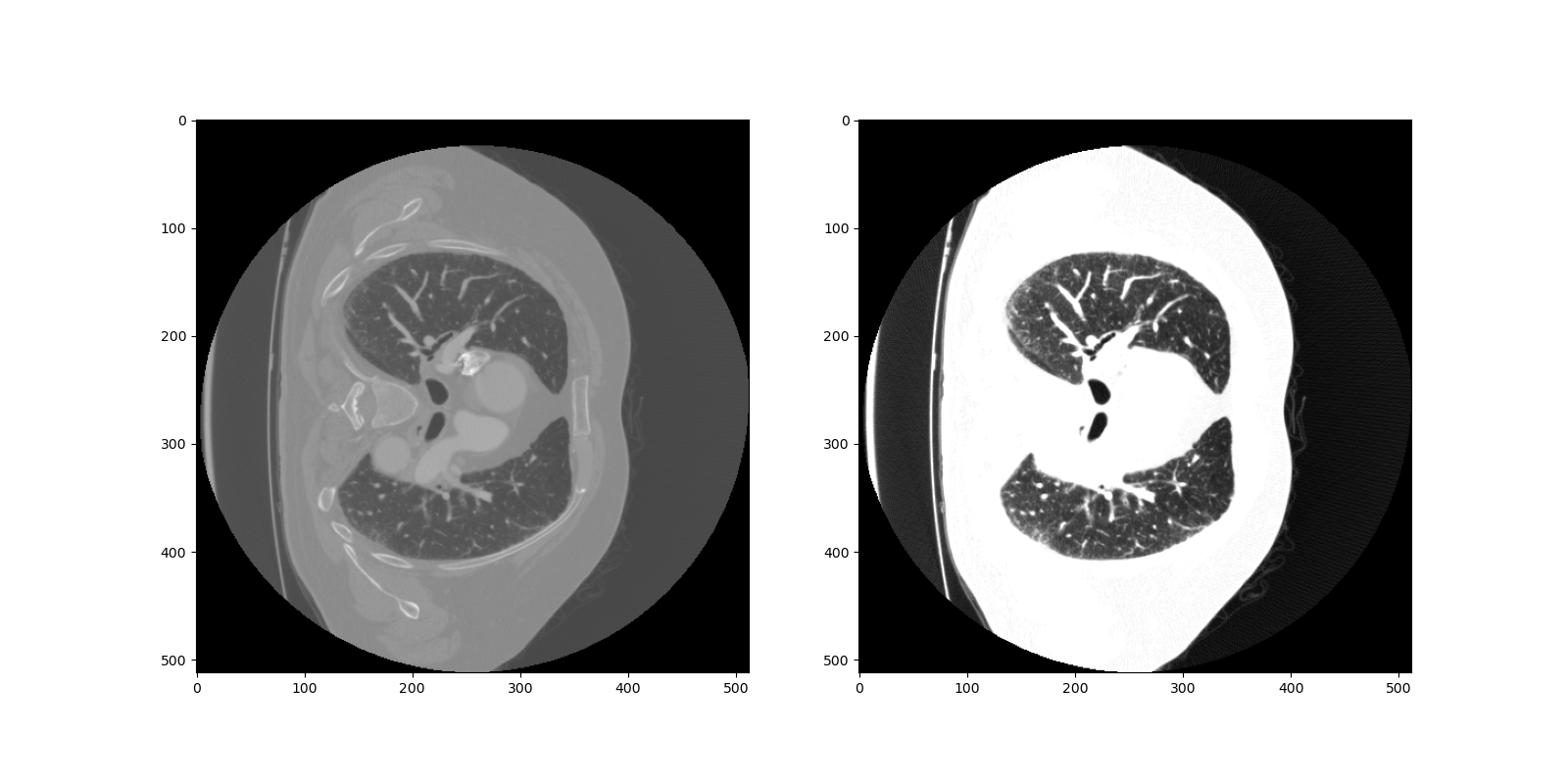}
\caption{Preprocessing of the inputs: The left one is the original slice and the right one is the input into the model.}\label{fig13}
\end{figure}

The training and test split in our model AGN is simple. Since the slices of CT scans are sequential, we do not need to shuffle the data to ensure the distributions of various cases in the slices in the training and testing are the similar. Hence, we just take the first 1500 slices as training part. Then, to feed the data into training, we have to do the preprocessing of the data slices. To be more specific, we have to adjust the level and window of the CT scans due to the fact that the Hounsfield ranges are different in different CT scans \cite{harris1993effect}. In other words, we have to rescale them to the same level. More importantly, we are able to enhance the contrast of the slices through the adjustment. The window and level in our case are 1000 and -600 respectively. Hence, we filter out the units out of the range (-1100, 100), which is illustrated in the Figure~\ref{fig13}.

The next step is the training process. The details of the training parameters are shown in the Table~\ref{tab2}.
\begin{table}[H]
\centering
\caption{Datasets details}\label{tab2}
\begin{tabular}{lll}
\hline
Dataset & Airway & CHASE\_DB1 \\
\hhline{===}
Vertex sampling sparsity $\delta$ & 4 & 4\\
Geodesic distance threshold $d$ &  10 & 40\\
CNN pretraining learning rate  & 0.01 & 0.001\\
Joint training learning rate &0.01&$5\times 10^{-3}$\\
Mini-batch size &1&1\\
CNN training iteration&150000&50000 \\
Joint training iteration&60000&50000\\
Graph update period&10000&10000\\
Negative slope in LeakyReLU&0.2&0.2\\
Dropout probability&0.1&0.1\\
\hline
\end{tabular}
\end{table}

In the pretraining CNN module and joint training process of AGN, the parameters we select are similar to those in the VGN. Note that the calculation burden is heavy if in each iteration we build the graph. To be more specific, since the time taken in the part of calculating the edges is long, which is almost several hours for each  iteration of training, it is acceptable that we build a new graph every 10000 iterations to save the training time.

Considering the number of iterations we select, we measure the loss as well as the dice coefficient during training. With a large number of iterations set, we run the model and reset a reasonable number of epoch for the total iterations when the model performance does not change so much to avoid overfitting. In addition, we select the number of iterations empirically, which is compared with CHASE\_DB1 dataset.

With respect to the loss functions and the optimizer, we utilise the binary cross-entropy loss and Adam optimizer. The binary cross-entropy loss function is advantageous particularly in calculating the probabilities, which is suitable in our case. We will take the mean of the loss function since we measure the overall performance of the model. Last but not least, we will utilise the dice coefficients to measure the similarities of prediction and the ground truth. The formula for the dice coefficients are given by $\text{dice coefficient}=\frac{2*(\text{prediction}\cap\text{ground truth})}{\text{prediction}+\text{ground truth}}$.

With the above configuration, we utilise the GPU at UCL cluster with memory 16 GB to train our AGN.

The details of all the related codes are stored on GitHub private repository.

The link is \url{https://github.com/ucapyyy/Airway-Graph-Network}.
\subsection{Results}
\begin{figure}[H]
\centering
\includegraphics[width=\textwidth]{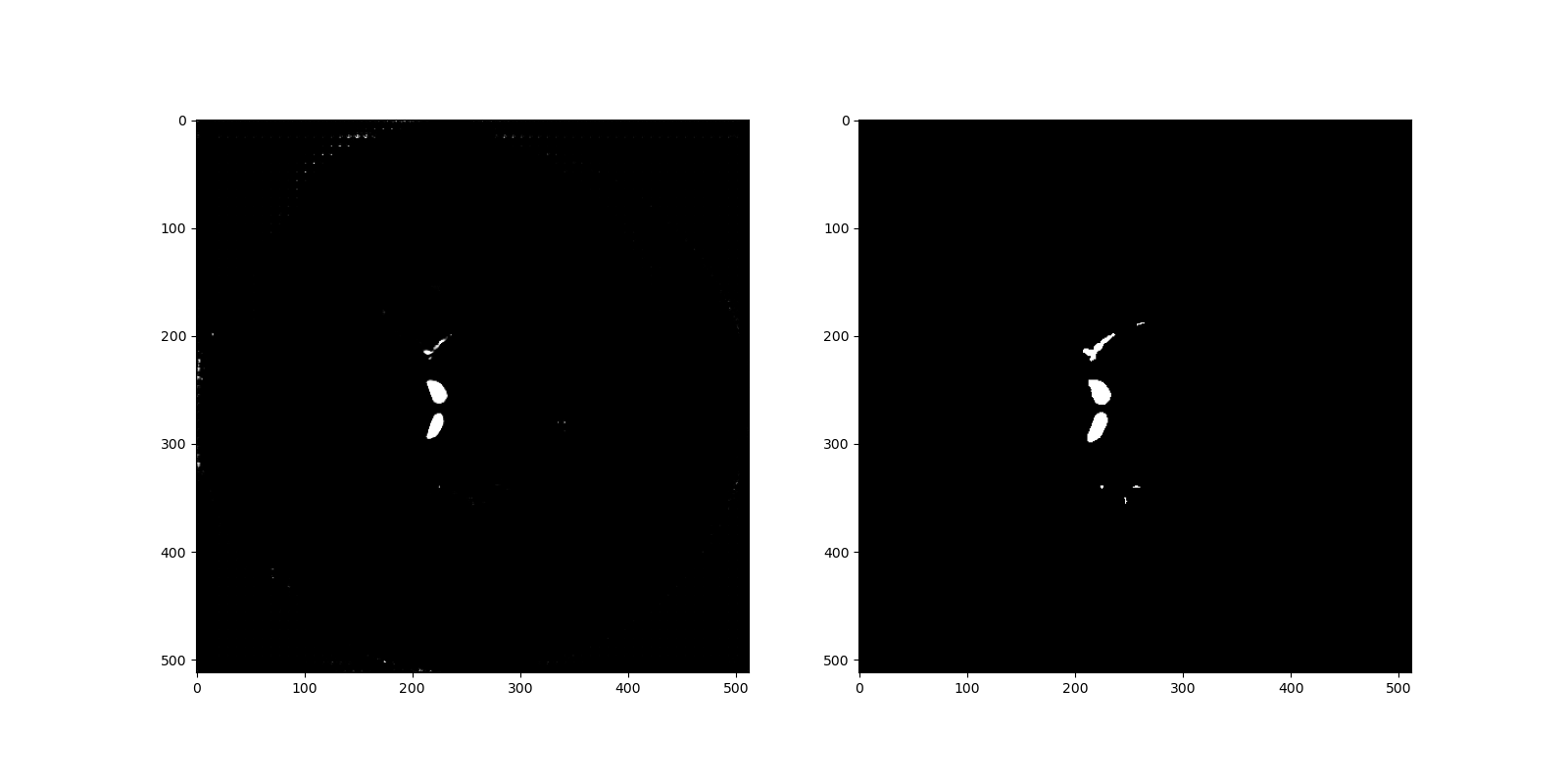}
\caption{Evaluation of the CNN pretrained module: The left one is the predicted segmentation and the right one is the segmentation of the ground truth.}\label{fig14}
\end{figure}

\begin{figure}[H]
\centering
\includegraphics[width=\textwidth]{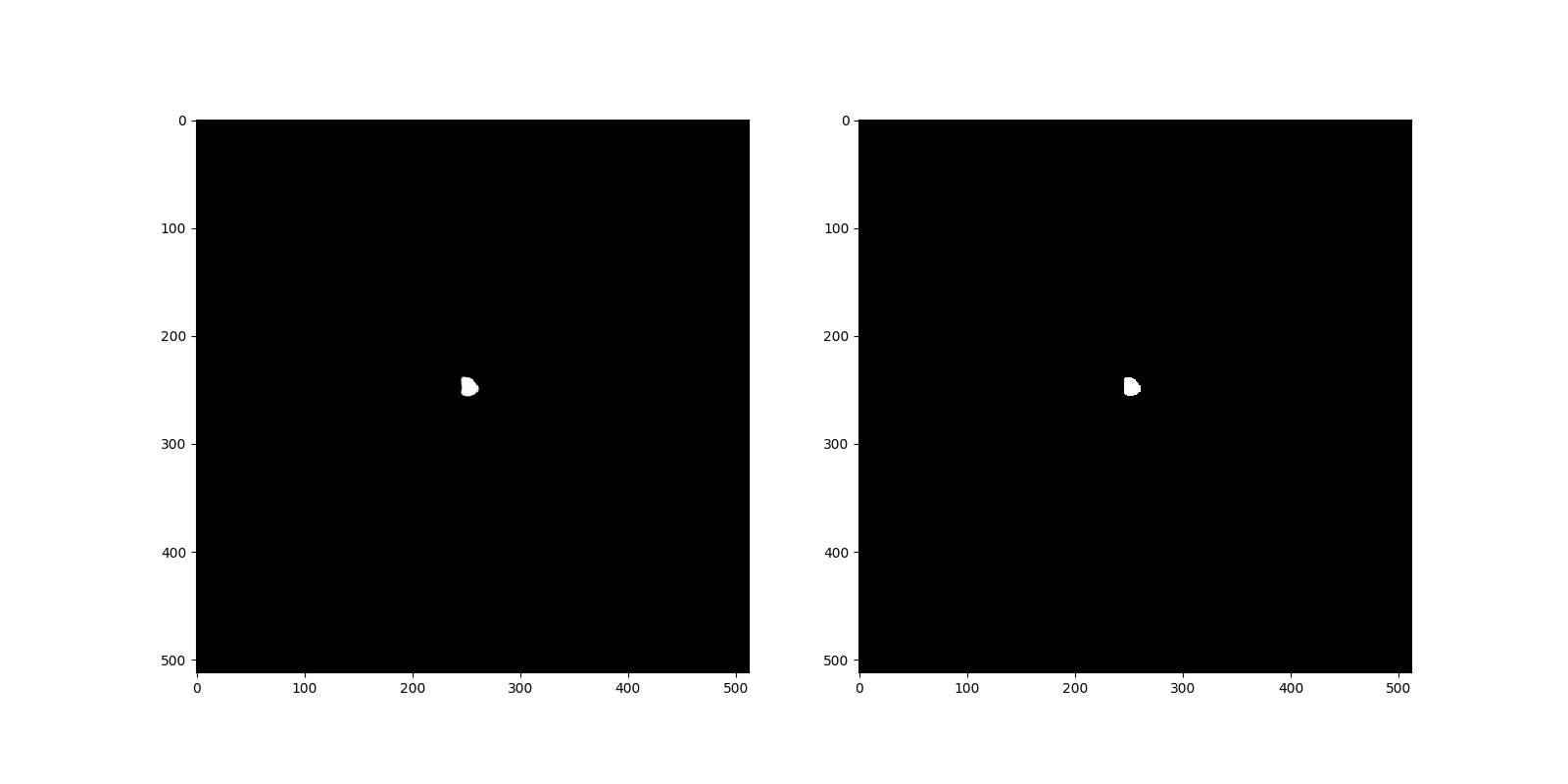}
\caption{Comparison of prediction and ground truth in some last several training iterations (only tracheae): The left one is the predicted segmentation and the right one is the segmentation of the ground truth}\label{fig15}
\end{figure}

In the Figure~\ref{fig14}, the segmentation from the CNN module in the test dataset is demonstrated. We can see from the two segmentations that the approximate segmentation is incomplete even if the tracheae are detected.
For the whole model training, we output the ground truth of the sampled slice as well as the predicted segmentation during training.
\newpage
In the Figure~\ref{fig15}, we can see that if the input slice only contains tracheae, then the prediction is good.
The dice coefficient for this segmentation is 0.9906, which means that the prediction and the ground truth are similar.

\begin{figure}[H]
\centering
\includegraphics[width=0.9\textwidth]{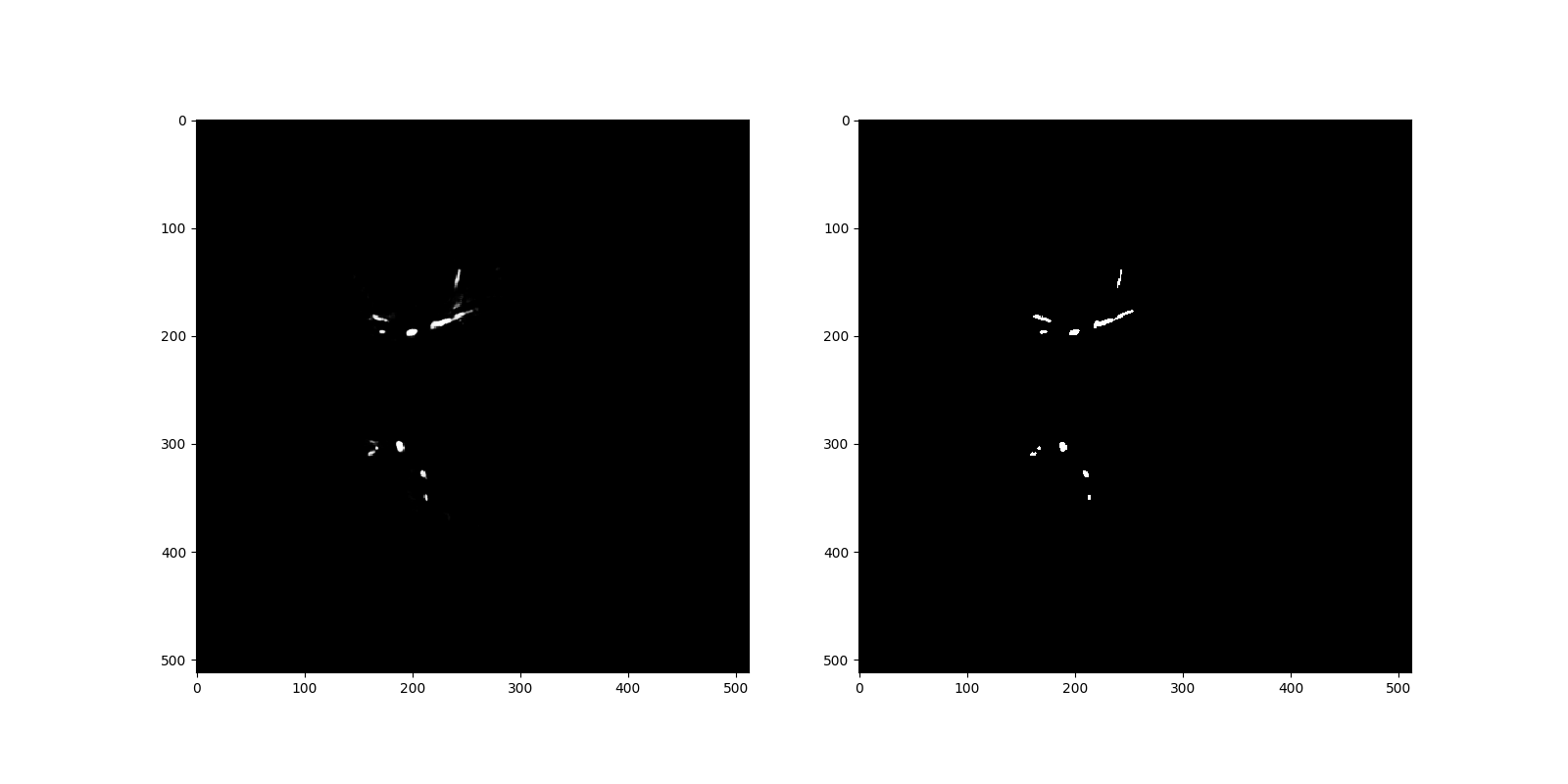}
\caption{Comparison of prediction and ground truth in some last several training iterations (including bronchi): The left one is the predicted segmentation and the right one is the segmentation of the ground truth.}\label{fig16}
\end{figure}

However, if the input slice contains bronchi, then there are false positive classifications.
In the Figure~\ref{fig16}, we can see that if the input slice contains bronchi, then the prediction is not so precise to some extent.
The dice coefficient for this segmentation is 0.8367, which means that the prediction and the ground truth are still similar but not as precise as that in the Figure~\ref{fig15}.

After the training process is completed, we compare the whole model  performance with the CNN-only pretrained model in the test dataset.
\begin{figure}[H]
\centering
\includegraphics[width=1.1\textwidth]{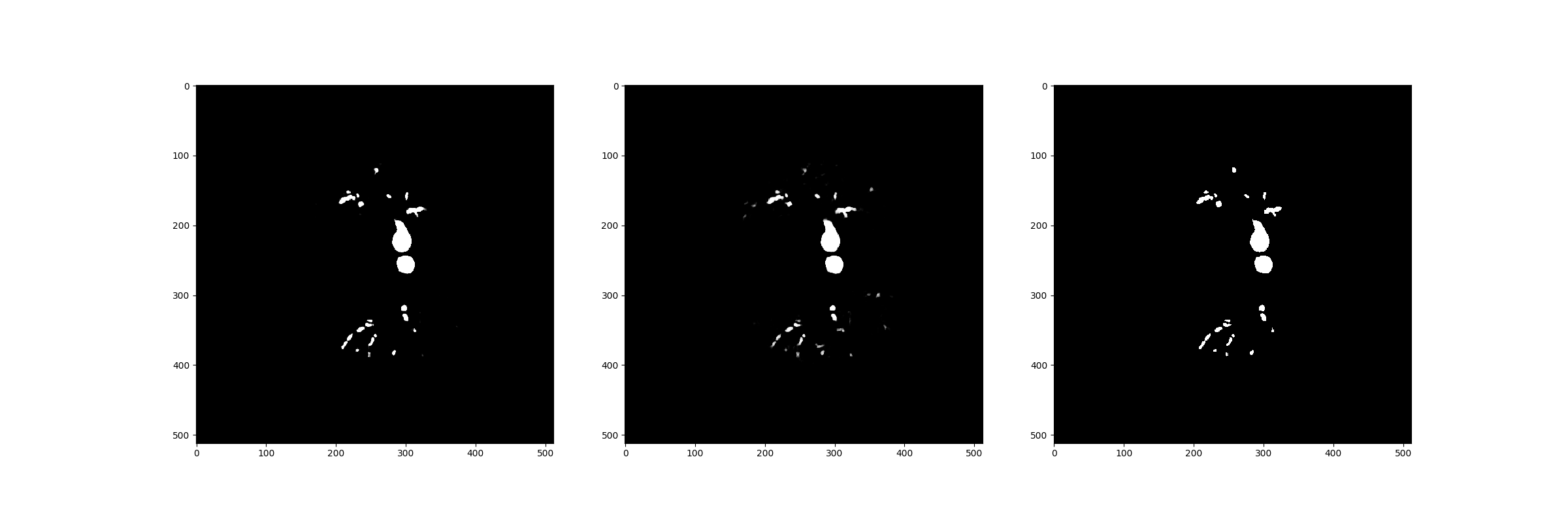}
\caption{Comparison of prediction of the whole model, CNN pretrained model and ground truth: The left one is the predicted segmentation of the whole model, the middle one is the counterpart of the CNN pretrained model and the right one is the segmentation of the ground truth.}\label{fig17}
\end{figure}
In the Figure~\ref{fig17}, we can find out that there are tiny changes of the bronchi that are detected in the whole model, which are improved by the GNN module.

Last but not least, we give the graph of the losses during training and the graph of dice coefficients during training.
\begin{figure}[H]
\centering
\includegraphics[width=0.6\textwidth]{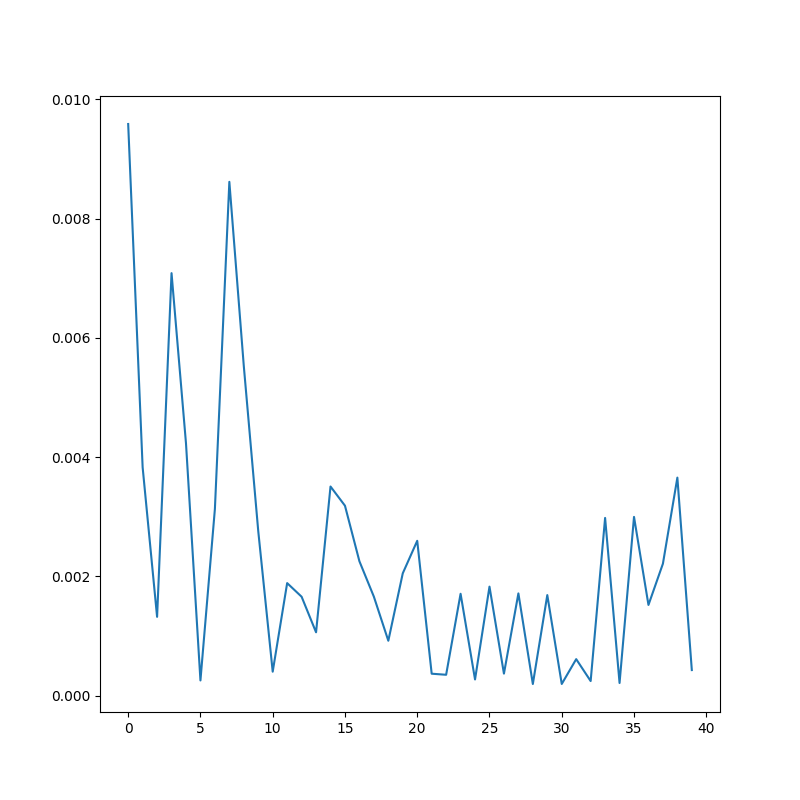}
\caption{Loss for training data during training}\label{fig18}
\end{figure}
In the Figure~\ref{fig18}, the binary cross-entropy losses of the training slices are shown. There is a decreasing trend of the loss during training. Hence, the final result of the model is valid.

As for the dice coefficients, since for the slices that contain little bronchi, which will cause dice coefficient to be 0, and the number of iterations are large, which will cause the plot to be too dense, we only plot the first epoch to see the model improvement during training.
\begin{figure}[H]
\centering
\includegraphics[width=0.6\textwidth]{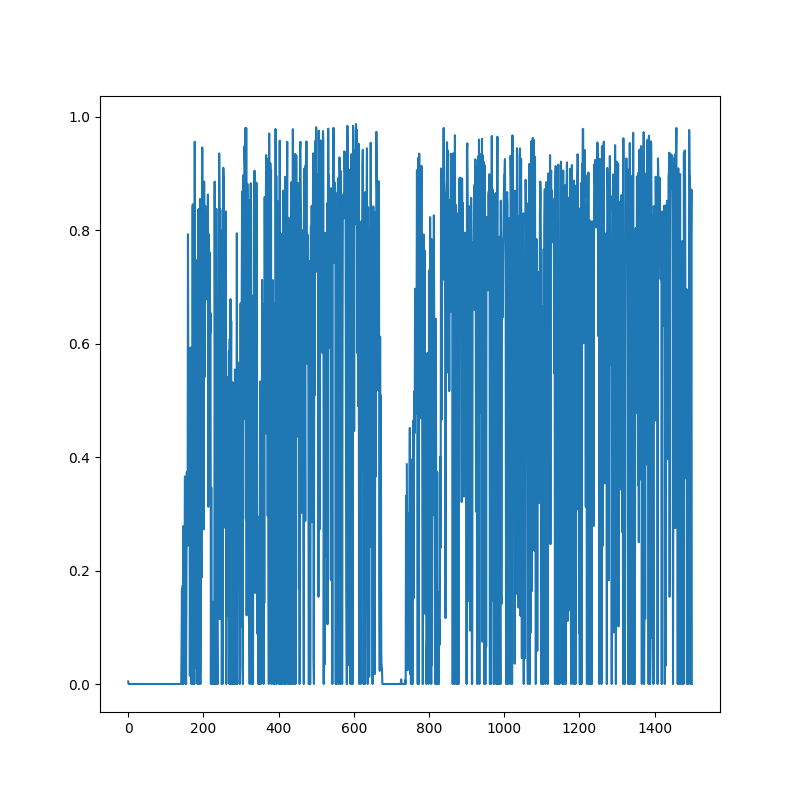}
\caption{Dice coefficient for training data during training in the first epoch}\label{fig19}
\end{figure}
In the Figure~\ref{fig19}, there are fluctuation of dice coefficients during training. However, we can see that in the first hundreds of iteration the dice coefficient is zero, meaning that the prediction is poor in the beginning. When the model learns the graph connectivity, the performance is better.

Not only do we need to plot the loss and dice coefficients for the training data, but also we have to plot the test data while training. We calculate the average of the loss and the dice coefficients respectively for all the test data every epoch.
\begin{figure}[H]
\centering
\includegraphics[width=0.6\textwidth]{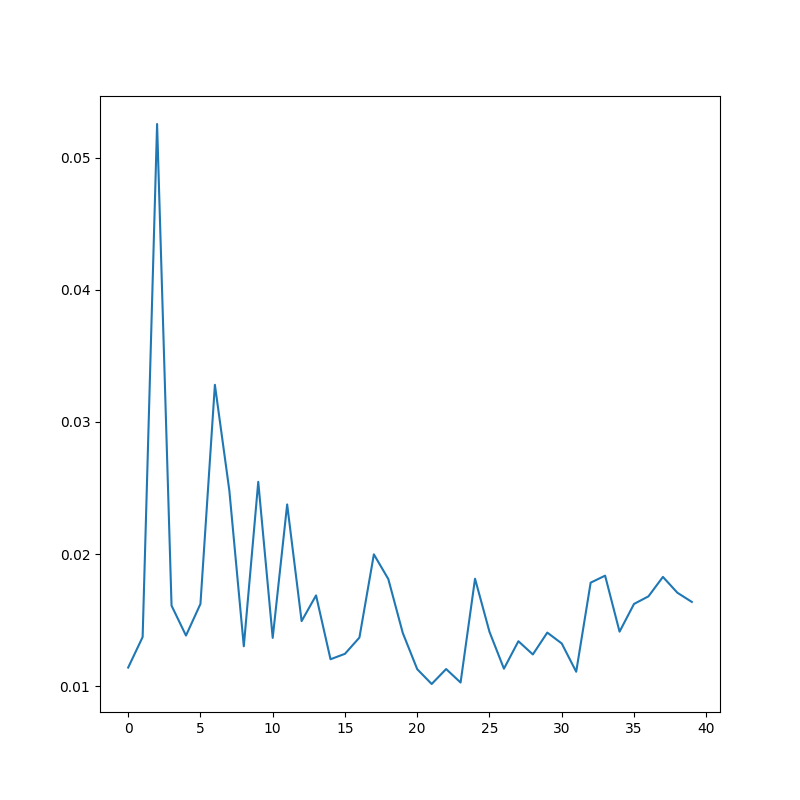}
\caption{Loss for test data during training}\label{fig20}
\end{figure}
In the Figure~\ref{fig20}, it shows a decreasing trend of loss in the test data.

\begin{figure}[H]
\centering
\includegraphics[width=0.6\textwidth]{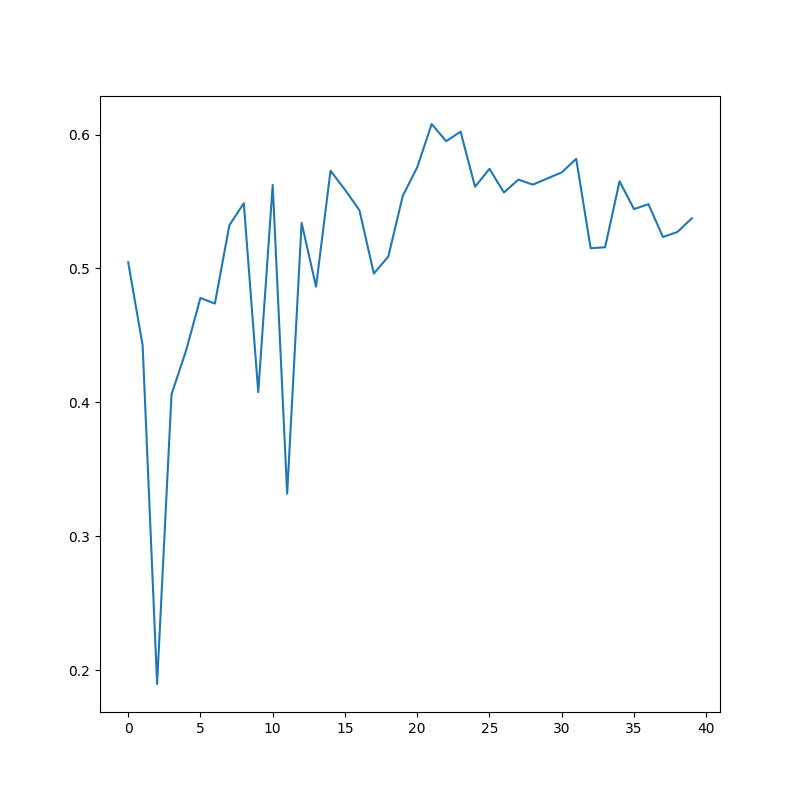}
\caption{Dice coefficient for test data during training}\label{fig21}
\end{figure}
For the dice coefficients, we have the similar results. In the Figure~\ref{fig21}, the corresponding dice coefficients for each epoch show an upward trend, meaning that the model performance is becoming better.
\newpage
\subsection{Discussion}
From the results, we are aware of the refinement when combining the GNN module. However, there are several potential problems in our model.

Firstly, the dataset in our model are axial slices of the CT scans. The shape of the tracheae in the slices are elliptical instead of line-shaped. Hence, the GNN module will not be able to build a satisfactory graph, i.e.\ the built graph may lose graph information. What is more, the bronchi in the slice are usually small dots, meaning that the nearby pixels do  not belong to the airways. It will not be extracted by the GNN module since there is no connectivity between it and the adjacent airway pixels. Actually, the adjacent airways are in the adjacent slices. The potential solution is to choose a different plane, namely the coronal plane. Moreover, the prediction in the CNN module is vital. In other words, without the approximately precise prediction in the bronchi, the whole model is possible to have a poor performance.

Secondly, the AGN is 2-dimensional. That is to say, the model will segment the airways in the 2D plane. When overlaying the slices of the predicted segmentation labels, the continuity between consecutive slices is likely to be lost. The potential solution is to post-process the predicted segmetation by checking the edges the segmentation.

Lastly, since the area occupied by the airways in the whole slice is small compared with the vessels in the retina images, the failure of the prediction in some bronchi is reasonable. The potential solution is to crop the slices and highlight the area of lungs by adding a pretrained model, where the model segments the approximate lungs in the whole slice.

\section{Conclusion}
In conclusion, the research project is aimed for the refinement of the CNN-only method in the work of airway segmentation. In the proposed AGN, with the CNN module considering local features and GNN module considering the global graph structures, we improve the performance of the segmentation from the CNN module owing to the fact that from the experiment, we measure the performances and find out the overall model outperforms the CNN module by comparing the segmentation in the CNN module and that in the overall model.  In addition, we plot the loss and the dice coefficient for the training data and the test data respectively for each each epoch to verify the model performance and determine the training parameters such as number of epochs. Last but not least, we summarise the potential weaknesses of the model and give possible explorations for future researches.

\newpage
\pagenumbering{gobble}
\bibliographystyle{splncs04}
\bibliography{references}
\end{document}